\documentclass[twocolumn,prl,aps,showpacs]{revtex4}

\usepackage{amsmath}
\usepackage{amssymb}
\usepackage[dvips]{graphicx}
\usepackage{psfrag}

\newcommand{\ra}{\rangle}
\newcommand{\la}{\langle}
\begin{document}

\title{Unifying All Classical Spin Models in a Lattice Gauge Theory}

\author{G.~De las Cuevas, W.~D\"ur, and H.~J.~Briegel}
\affiliation{
Institut f{\"u}r Theoretische Physik, Universit{\"a}t
Innsbruck, Technikerstra{\ss}e 25, A-6020 Innsbruck,
Austria\\
Institut f\"ur Quantenoptik und Quanteninformation der \"Osterreichischen Akademie der Wissenschaften, Innsbruck,Austria}

\author{M.~A.~Martin--Delgado}
\affiliation{
Departamento de F\'{\i}sica Te\'orica I, Universidad Complutense, 28040 Madrid, Spain}

\begin{abstract}
The partition function of all classical spin models, including all discrete standard statistical models and 
all Abelian discrete lattice gauge theories (LGTs), is expressed as a special instance of the partition 
function of the 4D  $\mathbb{Z}_2$ LGT. This uniÞes all classical spin models with apparently very different features in 
a single complete model. This result is applied to establish a new method to compute the mean-Þeld theory 
of Abelian discrete LGTs with $d 
\geq 4$, and to show that computing the partition function of the 4D  $\mathbb{Z}_2$ LGT 
is computationally hard ($\#$P hard). The 4D Z2 LGT is also proved to be approximately complete for 
Abelian continuous models. The proof uses techniques from quantum information.
\end{abstract}

\pacs{03.67.-a, 11.15.Ha, 03.67.Lx, 75.10.Hk, 05.50.+q}

\maketitle

\emph{1.~Introduction.---}Gauge theories describe the most fundamental interactions in nature, like
QED, weak interactions and QCD.
Only gravity has evaded a quantum version of the gauge principle. Lattice Gauge Theories (LGTs) are cutoff regulations of gauge theories of
strongly interacting particles~\cite{Ko79,Creutz83}.
When formulated on a lattice, LGTs become a new type of statistical mechanical models
 that are interesting by themselves, regardless of their connection with quantum gauge
theories in continuous space--time.

While Standard Statistical Models (SSMs) (like the classical Ising or Potts models~\cite{Wu84}) have global symmetries 
which are broken by local order parameters leading to different phases~\cite{It91}, LGTs have been introduced as models with local symmetries~\cite{Wegner71} which can only be broken by global order parameters, corresponding to closed string observables. LGTs have also been constructed for arbitrary 
non-Abelian gauge groups in the context of QCD to describe strong interactions~\cite{Wilson74}. 
Many interesting phenomena
are known to arise in LGTs which are different from SSMs. For example, LGTs may exhibit non--trivial phase diagrams that do not correspond to any continuum field theory. 
Already the simplest instances of LGTs ---Abelian discrete LGTs--- exhibit remarkable features such as a rich phase diagram depending on the gauge group $\mathbb{Z}_q$ and the underlying lattice, or confinement in the strong coupling limit.
In fact, it has been argued that the center of the group SU$(q)$, i.e.~the gauge group of Abelian discrete LGTs, $\mathbb{Z}_q$, plays an important role in the confinement problem~\cite{tHooft78}.

Given the variety of features of these models, a fundamental question arises: is it possible to give some structure to the set of all classical models? In this Letter we will give a positive answer to this question by showing that an LGT with $\mathbb{Z}_2$ gauge symmetry in a four dimensional square lattice, the 4D $\mathbb{Z}_2$ LGT, is complete with real coupling strengths for all Abelian discrete LGTs and all discrete SSMs. 
Here completeness means that the partition function of a large set of models (here, all classical spin models) can be expressed as a specific instance of the partition function of the complete model (thus, the notion of completeness can be seen as a form of universality). While similar completeness results for the 2D or 3D Ising model have been recently found in the context of SSMs~\cite{Va08,De08}, they are either restricted to specific classes of models~\cite{De08}, or are general but require complex coupling strengths, thereby lacking a physical interpretation~\cite{Va08}. 

The completeness results we present here are \emph{general} and are entirely based on \emph{real} coupling strengths, and thus are not only mathematical relations, but have physical implications. 
Note that all completeness results require to consider \emph{inhomogeneous} coupling strengths in the complete model. The results are general in the sense that they hold for all models with an arbitrary interaction pattern (including $k-$body interactions, and in any dimension $d$) between arbitrary $q-$level spins, which include all Abelian discrete LGTs and all discrete (Abelian) SSMs. The latter result establishes a general, explicit relation between models of very different physical origin (e.g.~models with local as opposed to global symmetries), in contrast with the specific and rather involved connections between certain SSMs and certain LGTs known before~\cite{Ko79}.
Furthermore, we will also show that the 4D $\mathbb{Z}_2$ LGT can efficiently approximate Abelian continuous LGTs with polynomial accuracy, as well as continuous SSMs, for which the efficiency depends on the scaling of the parameters, as will be specified below. To establish the proof, we use results from quantum information.

\emph{2.~Completeness of the 4D $\mathbb{Z}_2$ LGT.---}In order to prove the main result of this Letter, i.e.~the completeness of the 4D $\mathbb{Z}_2$ LGT, we first present a quantum formulation of the partition function of Abelian discrete LGTs (Sect.~2.1), which  allows us to construct systematic mappings to other models by properly choosing interaction strengths (Sect.~2.2). Then, we present a method to obtain general $n-$body interactions using only Ising--type interactions (Sect.~2.3). Finally, we shall use the manipulations of 2.2 to obtain all interactions required in 2.3 (Sect.~2.4).

\emph{2.1.~Quantum formulation of the partition function of Abelian discrete LGTs.---}We consider a standard definition of an Abelian discrete LGT
using a Wilson Hamilton function
in terms of face interactions, with gauge group $\mathbb{Z}_q$.
That is, we consider a lattice in $d$ dimensions in which each face $f\in F$
has $k$ sides or edges. Classical spins sit at the edges $e\in E$ of the
lattice, they have $q$ levels, $s_e=0,1,\ldots,q-1$, and they are subject
to a $k-$body interaction in every face:
\begin{equation}
H(\mathbf{s}) = - \sum_{f\in F} J_{1\ldots k} \cos\left(\frac{2\pi}{q}(s_{1}+s_{2}+\ldots+s_{k})\right),
\label{eq:H}
\end{equation}
where  $J_{1\ldots k}$ is the interaction strength at face $f$ that has spins $s_1, \ldots, s_k$ at its boundary, and the cosine depends on the sum of the spins modulo $q$.  For $q=2$ (and any $k$ and $d$) we shall refer to these interactions as ``Ising--type interactions'', and to the model as ``$d$ $\mathbb{Z}_2$ LGT'' (by default meaning $k=4$). Note that in this case each term in the Hamilton function takes the form $J_{1\ldots k}(-1)^{s_1+\ldots+s_k}$, thereby only depending on the parity of the $k$ adjacent spins.
The Hamilton function~\eqref{eq:H} is invariant under the local transformation
$g_v=\prod_{e:e\textrm{ adj }v} X_e$,
where $e\textrm{ adj }v$ are edges adjacent to a vertex $v$, and $X_e$ is defined as $X_e: s_e \to (s_e+1)_{\textrm{mod }q}$. The gauge group is generated by these transformations: $ \mathbb{Z}_q= \langle g_v, \forall v\in V \rangle $.
The partition function of this system is defined as
\begin{equation}
Z_{\mathrm{LGT}}=\sum_{\mathbf{s}}e^{-\beta H(\mathbf{s})},
\label{eq:Z}
\end{equation}
where $\beta = 1/(k_B T)$ is the inverse temperature, and $\mathbf{s}:=(s_1,\ldots,s_{|E|})$ is the spin configuration.

\begin{figure}[t]
\centering
\psfrag{1}{$s_1$}
\psfrag{2}{$s_2$}
\psfrag{3}{$s_3$}
\psfrag{4}{$s_4$}
\psfrag{5}{$s_5$}
\psfrag{6}{$s_6$}
\psfrag{7}{$s_7$}
\psfrag{8}{$s_8$}
\psfrag{a}{$J_{1234}$}
\psfrag{b}{$J_{4567}$}
\psfrag{c}{$J_{4567}$}
\psfrag{g}{}
\psfrag{h}{}
\psfrag{i}{}
\psfrag{j}{}
\psfrag{A}{(a)}
\psfrag{B}{(b)}
\includegraphics[width=1\columnwidth]{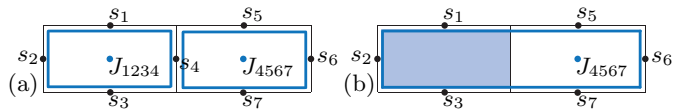}
\caption{(color online) (a) The state $|\psi_{\mathrm{LGT}}\ra$ (here, on a 2D square lattice) places one quantum particle (blue dots) at every face, which characterizes the interaction of classical spins (black dots) on that face. (b) The left face is merged with the right one by setting $J_{1234}=\infty$ (marked with a shaded face).}
\label{fig:1}
\end{figure}

Now we present a quantum formulation of \eqref{eq:Z}. In the same spirit as in \cite{Va08,Va07}, we define a quantum state of $|F|$ $q-$level quantum particles, which we can imagine to sit at the center of each face, as in Fig.~\ref{fig:1}(a),
\begin{eqnarray}
|\psi_\mathrm{LGT}\ra &=& \sum_{\mathbf{s}} \bigotimes_{f\in F} | s_1+\ldots+s_k\ra
\label{eq:psiLGT}
\end{eqnarray}
where $s_1,\ldots,s_k$ are the (values of the) spins at the boundary of face $f$ and they are summed modulo $q$.
We also define a product state $|\alpha\ra = \bigotimes_{f\in F} |\alpha_f\ra$ with
\begin{equation}
|\alpha_f\ra = \sum_{s_1,\ldots,s_k} e^{\beta J_{1\ldots k}  \cos\left[\frac{2\pi}{q}(s_1+\ldots+s_k)\right]} |s_1+\ldots +s_k\ra .
\label{eq:alpha_f}
\end{equation}
The basis states in \eqref{eq:psiLGT} and \eqref{eq:alpha_f} are the eigenstates of the quantum phase shift operator $Z|j\ra = e^{i2\pi j/q}|j\ra$, for $j=0,1,\ldots,q-1$.
The partition function of the Abelian discrete LGT \eqref{eq:Z} is then obtained by
computing the overlap between $|\psi_\mathrm{LGT}\ra$ and $| \alpha \ra$:
\begin{equation}
Z_\mathrm{LGT} = \la \alpha | \psi_\mathrm{LGT}\ra .
\label{eq:ZLGT}
\end{equation}

\emph{2.2.~Merge and deletion rules.---}Similar as in~\cite{Va08,De08} we now show how to manipulate the state $|\psi_{\mathrm{LGT}}\ra$ by means of measurements. That is, we show how to choose a product state $|\gamma\rangle = \otimes _{f \in \tilde F}|\alpha_f\rangle$ on a subset $\tilde F$ of quantum particles such that the state of the remaining particles after the projection onto $\langle \gamma|$, $ |\psi'_{\mathrm{LGT}}\ra=(\la\gamma| \otimes I) |\psi_{\mathrm{LGT}}\ra$,  is again of the form \eqref{eq:psiLGT}, but now defined on a different lattice $L'$. This allows us to relate the partition function of models defined on $L'$ to the partition function of the 4D $\mathbb{Z}_2$ LGT via~\eqref{eq:ZLGT}.
More precisely, we consider the \emph{merge rule of faces}, which corresponds to setting $|\alpha_{f'}\ra=|0_{f'}\ra$ for a certain face $f'$. 
From~\eqref{eq:alpha_f} we see that this choice corresponds to setting $J_{f'}=\infty$~\cite{Va08}. 
Moreover, applying this projection on state~\eqref{eq:psiLGT} we see that it sets the condition $s_1+\ldots +s_k=0$ for the $k$ spins at the boundary of $f'$. Now we can substitute this condition on a neighboring face which, instead of depending on, say, $s_k$, will now depend on $s_1+\ldots+s_{k-1}$. For example, if we apply $\la0|$ on the left face of Fig.~\ref{fig:1}(a), we impose the condition $s_1+s_2+s_3+s_4=0$. This can be substituted in the right face, which will now depend on $(s_1+s_2+s_3)+s_5+s_6+s_7$, thereby effectively \emph{merging} the two faces into one larger face with an Ising--type interaction with strength $J_{4567}$ (Fig.~\ref{fig:1}(b)).
We note that the merge rule can be concatenated, thereby enlarging faces at will.
On the other hand, the \emph{deletion rule} of faces corresponds to projecting a face $f'$ onto the symmetric state $\sum_{s_1,\ldots,s_k}\la s_1+\ldots+s_k|$, i.e.~to setting $J_{f'}=0$. 

\emph{2.3.~Method to obtain arbitrary $n-$body interactions.---}Next we show that we can generate a totally general interaction between $n$ 2$-$level particles if all Ising--type $k-$body interactions between these $n$ particles are available, for any subset of $k$ particles and all $k=0,1,\ldots,n$. 

A general interaction between $n$ spins corresponds to assigning a different energy $\lambda_{\mathbf{s}}$ to each spin configuration $\mathbf{s}$. Hence, we need to show that there always exists a combination of Ising--type interactions on the different subsets, i.e.~some $J, J_1,\ldots, J_n,J_{12},\ldots,J_{1\ldots n}$ (the subindex indicates which particles participate in that interaction) such that $J(-1)^{0} + J_1(-1)^{s_1}+\ldots+J_n(-1)^{s_n}+J_{12}(-1)^{s_1+s_2}+\ldots + J_{12\ldots n}(-1)^{s_1+\ldots+s_n} = \lambda_{\mathbf{s}}$ is satisfied for arbitrary $\lambda_{\mathbf{s}}$ and for all $\mathbf{s}$. This is equivalent to showing that a matrix with one prefactor of the $J$'s in each column (i.e.~$(-1)^{0}, (-1)^{s_1},\ldots, (-1)^{s_n}, \ldots$) and with one row for each spin configuration is always invertible. Notice that this is a square matrix, since there are 
$\sum_{k=0}^{n}\binom{n}{k}=2^n$
 $J$'s (columns), and $2^n$ $\lambda_\mathbf{s}$'s (rows). But by construction all rows are linearly independent, hence it has nonzero determinant, and thus it is invertible.

\emph{2.4.~Explicit construction.---}We now use the tools of 2.2 in order to show that we can obtain all interactions required in 2.3 in a 4D $\mathbb{Z}_2$ LGT. The proof will require to fix some spins using the gauge symmetry of the model (i.e.~fixing them to zero while leaving the Hamilton function invariant), a technique which can be applied as long as the edges whose spins are fixed form at most a maximal tree (i.e.~do not form a closed loop)~\cite{maxtree}.

\begin{figure}[t]\centering
\psfrag{3}{$s_1$}
\psfrag{1}{$s_1$}
\psfrag{2}{$s_2$}
\psfrag{a}{$s_1$}
\psfrag{b}{$s_2$}
\psfrag{c}{$s_3$}
\psfrag{l}{$r$}
\psfrag{r}{$r_1$}
\psfrag{p}{$r_2$}
\psfrag{A}{(a)}
\psfrag{B}{(b)}
\psfrag{C}{(c)}
\psfrag{s}{$J_1$}
\psfrag{u}{$J_{12}$}
\psfrag{t}{$J_{123}$}
\includegraphics[width=0.95\columnwidth]{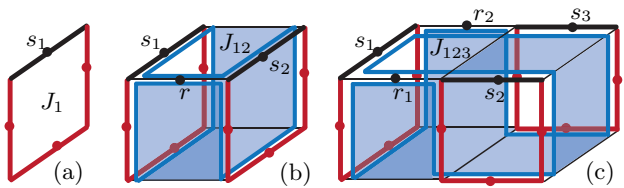}
\caption{(color online) Spins fixed by the gauge are marked in red. A single--body, a two--body and a three-body Ising--type interaction with coupling strengths $J_1$, $J_{12}$ and $J_{123}$ are shown in (a), (b) and (c), respectively.
}
\label{fig:2}
\end{figure}

First, a ``single--body interaction'' of $s_1$ (analogous to a magnetic field) is obtained by letting $s_1$ interact with all other spins around a face fixed by the gauge (Fig.~\ref{fig:2}(a)).

A two--body interaction is obtained by merging the front, lower and back face
and creating the face with blue boundaries of Fig.~\ref{fig:2}(b). By fixing with the gauge six of the spins at the boundary of the blue face, this face depends only on $s_1+s_2+r+r=s_1+s_2$ (since the sum is mod 2) (see Fig.~\ref{fig:2}(b)). Thus, this effectively corresponds to a two--body Ising--type interaction between $s_1$ and $s_2$ with an interaction strength $J_{12}$. Notice that by setting $J_{12}=\infty$ as well, we force $s_1+s_2=0$, which can be seen as a propagation of $s_1$ into $s_2$ (since $s_1=s_2$). A concatenated application of this two--body interaction results in an effective propagation of a spin through a certain path (the turnings of the path can be done similarly).

A three--body interaction is obtained by bringing three spins $s_1,s_2,s_3$ close to each other and then merging the large blue face as indicated in Fig.~\ref{fig:2}(c). The interaction in the blue face corresponds to a three--body Ising--type interaction between $s_1,s_2$ and $s_3$ (since $r_1$ and $r_2$ are summed twice) and with an interaction strength $J_{123}$.

The generalization to $k-$body interactions with $k\geq 4$ can be done in a similar way as in Fig.~\ref{fig:2}(c). The spins $s_j$ taking place in the final $k-$body interaction are never adjacent, and each of them is part of a face at the front, back or side with three spins fixed by the gauge (red u--shapes). All but one of the remaining faces are merged, and the interaction strength $J_{1\ldots k}$  in that face determines the $k-$body Ising--type interaction.

\begin{figure}
\psfrag{a}{{\small $s_3$}}
\psfrag{b}{{\small $s_2$}}
\psfrag{c}{{\small $s_1$}}
\psfrag{d}{{\small $s_4$}}
\psfrag{f}{{\small $s_5$}}
\psfrag{g}{{\small $s_4$}}
\psfrag{v}{{\small $s_6$}}
\psfrag{s}{{\small $r_1$}}
\psfrag{r}{{\small $r_2$}}
\psfrag{t}{{\small $r_3$}}
\psfrag{u}{{\small $r_4$}}
\psfrag{w}{{\small $r_5$}}
\includegraphics[width=0.77\columnwidth]{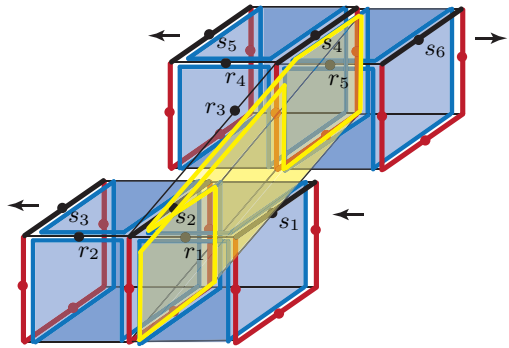}
\caption{(color online) Replication of spins in four dimensions. Yellow faces represent the fourth dimension, and they have the same meaning as blue faces, that is, $s_2$ propagates into $s_3$ by the same method as it propagates into $s_4$.}
\label{fig:3}
\end{figure}

Thus we have shown how to obtain $k-$body Ising--type interactions between any group of $k$ particles, for $k=1,\ldots, n$ (the zero--body interaction required in 2.3 is a constant factor, so we obtain $Z$ up to this factor). Since the total number of interactions is $2^n$, we only need to show that a given spin can participate in $2^n/n$ interactions. This means that each spin must have this number of ``end faces'', i.e.~faces at the end of a propagation that participate in a (many--body) interaction. For example, if we use Fig.~\ref{fig:2}(b) to propagate $s_1$ (i.e.~we set $J_{12}=\infty$), then $s_1$ has two end faces, the left and the right one, each of which can participate in, say, a three--body interaction like the one shown in Fig.~\ref{fig:2}(c). But, as can be seen from Fig.~\ref{fig:2}(b), the propagation of a particle (in 3D) essentially behaves as a ``pipe'' which has only two end faces. In fact, the number of ends that an encoded particle of dimension  $d_e$ in a lattice of dimension $d$ can have are $2(d-d_e)$. Here we essentially have $d_e=2$, and thus for $d=3$ the particle is blocked to have only 2 ends. We need to resort to a 4D lattice in order to obtain $2(d-d_e)>2$ ends, and then this replication in different directions can be multiply applied until the particle has $2^n/n$ ends (see Fig.~\ref{fig:3} for a replication of one spin $s_1$ into three other ``end faces'' $s_3$, $s_5$ and $s_6$).
We refer the reader to \cite{De08a} for the detailed construction. 
We remark that all faces which are not mentioned in this construction have to be deleted using the deletion rule.
We also mention that we have tried several other procedures to obtain this result in 3D, but none of them could avoid the formation of loops of edges fixed by the gauge.

This proves that we can generate a totally general $n-$body interaction between $n$ particles in a 4D $\mathbb{Z}_2$ LGT. This includes all classical spin models with $q=2$ in arbitrary dimensions $d$, arbitrary graphs, and arbitrary interaction pattern. Moreover, by  encoding a $q-$level particle in $m_q=\lceil \log q \rceil$ $2-$level particles, this also includes general interactions between $n'$ $q-$level particles, with $n'=n/m_q$. This proves that the 4D $\mathbb{Z}_2$ LGT is complete for all Abelian discrete classical spin models, including all Abelian discrete LGTs and discrete SSMs.

\emph{2.5.~Approximate completeness for Abelian continuous LGTs and continuous SSMs.---}We can go further and show that the 4D $\mathbb{Z}_2$ LGT is also approximately complete for Abelian continuous models, that is, the partition function of a continuous model can be expressed, up to a certain accuracy, as a specific instance of the partition function of the 4D $\mathbb{Z}_2$ LGT. To see this, we just need to let $q\to\infty$ (the lattice spacing remaining discrete) and determine what approximation can be obtained (see below).

\emph{2.6.~Efficiency results.---}The construction presented above enables one to generate, from a 4D $\mathbb{Z}_2$ LGT, Hamilton functions that contain $M$ terms with at most $k-$body interactions with an overhead that scales poly($M,2^k$) for $q=2$. 
In the case of $q-$state models with $M$ general $k'-$body interaction terms, at most $2^{k'm_q}$ Ising--type interactions between $k'm_q$ $2-$level particles are required for each term. Therefore, the overhead in the system size of the 4D $\mathbb{Z}_2$ LGT w.r.t the final model is polynomial if $k'$ scales not faster than logarithmically, and $q$ and $M$ scale polynomially with the system size. These criteria determine whether a given continuous SSM can be approximated efficiently. Abelian continuous LGTs usually have $k$ fixed (e.g. $k=4$), and thus they can be efficiently approximated by letting $q \to\infty$ polynomially.

\emph{3.~Implications of the main result.---}We shall now draw three conclusions from the main result.

\emph{3.1.~Completeness of the 3D $\mathbb{Z}_2$ LGT with fixed boundary conditions.---}As mentioned above, the only obstacle in proving that the 3D $\mathbb{Z}_2$ LGT is complete was that the replication of spins combined with the $k-$body interactions caused loops of spins fixed by the gauge. This problem can be overcome by allowing for fixed boundary conditions, i.e.~fixing spins to zero at the boundary of the 3D lattice. This means that the completeness results for the 4D $\mathbb{Z}_2$ LGT also hold for the 3D $\mathbb{Z}_2$ LGT with fixed boundary conditions, which has the same bulk interactions in the thermodynamic limit as the 3D $\mathbb{Z}_2$ LGT.

\emph{3.2.~Mean--field theory for $\mathbb{Z}_2$ LGTs at fixed $d\geq 4$.---}
While the mean--field theories of SSMs are easy to construct, this is
not the case for LGTs since Elitzur's theorem~\cite{El75} prevents the non--vanishing
mean value of a link variable.
This problem was circumvented using a saddle--point approximation~\cite{Drouffe81,FLZ82} with the
inverse dimension $1/d$ as an expansion parameter. The restoration of
the gauge symmetry is nontrivial in this expansion.
Here we propose a new method that does not break gauge invariance and that works for fixed $d$, with $d\geq 4$, which is based on the construction of a $k-$clique, i.e.~a graph with all possible (here, Ising--type) $k-$body interactions. Thus, constructing such a graph is a way of averaging over the interaction of a given particle with all the rest, and, hence, a way of computing its mean--field theory. The construction of the 4--clique for the 4D $\mathbb{Z}_2$ LGT has been shown in Sec.~2.4 (where we constructed the $k-$ cliques for all $k=1,\ldots,n$). The same construction applies trivially to $d$ $\mathbb{Z}_2$ LGTs with $d> 4$, simply by not using the extra dimensions (see~\cite{De08a} for further details).
We mention that this method has potential applications in computer simulations.

\emph{3.3.~Computational complexity of the 4D $\mathbb{Z}_2$ LGT.---}
Our results imply, in particular, that the partition function of the 2D Ising model with magnetic fields can be expressed as a specific instance of the partition function of the 4D $\mathbb{Z}_2$ LGT. Because the former is known to be \#P complete problem~\cite{Ba82,Aa},  we conclude that the problem of evaluating the partition function of the 4D $\mathbb{Z}_2$ LGT in the real parameter regime is \#P hard, i.e.~computationally difficult.
On the other hand, one can show that approximating the partition function of the 3D and 4D $\mathbb{Z}_2$ LGT in a certain complex parameter regime with polynomial accuracy is as hard as simulating arbitrary quantum computations, i.e.~BQP complete~\cite{De08a}.

\emph{4.~Outlook.---} Our results provide a unification of all classical models with very different features into a single complete model, the 4D $\mathbb{Z}_2$ LGT. In particular, models with different types of order parameters, as well as models belonging to different universality classes can be obtained. It will be interesting to use our results to gain further insight in the structure of classical spin models.

We thank M.~Van den Nest for valuable discussions and for pointing out the connections to complexity. Work supported by the FWF,  the European Union (QICS, SCALA),
FIS2006-04885, and ESF INSTANS 2005-10.


\begin{thebibliography}{99}

\bibitem{Ko79}
J.~B.~Kogut,
%\emph{An introduction to lattice gauge theories and spin systems.}
Rev.~Mod.~Phys.~\textbf{51}, 659 (1979).

\bibitem{Creutz83}
M.~Creutz, \emph{Quarks, gluons and lattices},
Cambridge University Press (1983).

\bibitem{Wu84}
F.~Y.~Wu, Rev.~Mod.~Phys.~{\bf 54}, 235 (1982).

\bibitem{It91}
C.~Itzykson, J.~M.~Drouffe,~\emph{Statistical~Field~Theory} (Cambridge University Press, Cambridge, England, 1991).

\bibitem{Wegner71}
F.~Wegner, J.~Math.~Phys.~(N.Y.) {\bf 12}, 2259 (1971).

\bibitem{Wilson74}
K.~G.~Wilson, Phys.~Rev.~D {\bf10}, 2445 (1974).

\bibitem{tHooft78}
G.~H.~'t Hooft, Nucl.~Phys.~{\bf B138}, 1 (1978).

\bibitem{De08}
G.~De las Cuevas \emph{et al.} arXiv:0812.2368, [JSTAT (to be published)]

\bibitem{Va08}
M.~Van den Nest, W.~D\"ur and H.J.~Briegel,
Phys.~Rev.~Lett.~{\bf 100}, 110501 (2008).

\bibitem{Va07}
M.~Van den Nest, W.~D\"ur and H.J.~Briegel,
Phys.~Rev.~Lett.~{\bf 98}, 117207 (2007).

\bibitem{maxtree}
M.~Creutz, Phys.~Rev.~D {\bf 15}, 1128 (1977).


\bibitem{De08a}
G.~De las Cuevas \emph{et al.} (to be published).

\bibitem{El75}
S.~Elitzur,
%\emph{Impossibility of spontaneously breaking local symmetries.}
Phys.~Rev.~D \textbf{12}, 3978 (1975).

\bibitem{Drouffe81}
J.~M.~Drouffe, Phys. Lett. B {\bf 105}, 46 (1981).

\bibitem{FLZ82}
H.~Flyvberg \emph{et al.}, %B. Lautrup, J.B. Zuber,
Phys.~Lett.~B{\bf110}, 279 (1982).

\bibitem{Ba82}
F.~Barahona, J.~Phys.~A {\bf 15}, 3241 (1982).

\bibitem{Aa}
S.~Aaronson, The complexity zoo. \\
\verb|http://qwiki.stanford.edu/wiki/Complexity_Zoo|

\end{thebibliography}
\end{document}